
%
\def\unlockat{\catcode`\@=11}
\def\lockat{\catcode`\@=12}
\unlockat
\def\d@f@ult{} \newif\ifamsfonts \newif\ifafour
\def\m@ssage{\immediate\write16}  \m@ssage{}
\m@ssage{hep-th preprint macros.  Last modified 16/10/92 (jmf).}
\message{These macros work with AMS Fonts 2.1 (available via ftp from}
\message{e-math.ams.com).  If you have them simply hit "return"; if}
\message{you don't, type "n" now: }
\endlinechar=-1  
\read-1 to\@nswer
\endlinechar=13
\ifx\@nswer\d@f@ult\amsfontstrue
    \m@ssage{(Will load AMS fonts.)}
\else\amsfontsfalse\m@ssage{(Won't load AMS fonts.)}\fi
\message{The default papersize is A4.  If you use US 8.5" x 11"}
\message{type an "a" now, else just hit "return": }
\endlinechar=-1  
\read-1 to\@nswer
\endlinechar=13
\ifx\@nswer\d@f@ult\afourtrue
    \m@ssage{(Using A4 paper.)}
\else\afourfalse\m@ssage{(Using US 8.5" x 11".)}\fi
\nonstopmode
%
%

\font\twelverm=cmr12
\font\ninerm=cmr9
\font\sixrm=cmr6
\font\fourteenbf=cmbx12 scaled\magstep1
\font\twelvebf=cmbx12
\font\ninebf=cmbx9
\font\sixbf=cmbx6
\font\fourteeni=cmmi12 scaled\magstep1      \skewchar\fourteeni='177
\font\twelvei=cmmi12                        \skewchar\twelvei='177
\font\ninei=cmmi9                           \skewchar\ninei='177
\font\sixi=cmmi6                            \skewchar\sixi='177
\font\fourteensy=cmsy10 scaled\magstep2     \skewchar\fourteensy='60
\font\twelvesy=cmsy10 scaled\magstep1       \skewchar\twelvesy='60
\font\ninesy=cmsy9                          \skewchar\ninesy='60
\font\sixsy=cmsy6                           \skewchar\sixsy='60
\font\fourteenex=cmex10 scaled\magstep2
\font\twelveex=cmex10 scaled\magstep1

\ifamsfonts
   \font\ninex=cmex9
   
   \font\sixex=cmex7 at 6pt
   
\else
   \font\ninex=cmex10 at 9pt
   
   \font\sixex=cmex10 at 6pt
   
\fi
\font\fourteensl=cmsl10 scaled\magstep2
\font\twelvesl=cmsl10 scaled\magstep1

\font\sevensl=cmsl10 at 7pt
\font\sixsl=cmsl10 at 6pt

\font\fourteenit=cmti12 scaled\magstep1
\font\twelveit=cmti12

\font\fourteentt=cmtt12 scaled\magstep1
\font\twelvett=cmtt12
\font\fourteencp=cmcsc10 scaled\magstep2
\font\twelvecp=cmcsc10 scaled\magstep1

\ifamsfonts
   
\else
   
\fi
\newfam\cpfam
\font\fourteenss=cmss12 scaled\magstep1
\font\twelvess=cmss12
\font\tenss=cmss10
\font\niness=cmss9

\font\sevenss=cmss8 at 7pt
\font\sixss=cmss8 at 6pt
\newfam\ssfam
\newfam\msafam \newfam\msbfam \newfam\eufam
\ifamsfonts
 \font\fourteenmsa=msam10 scaled\magstep2
 \font\twelvemsa=msam10 scaled\magstep1
 \font\tenmsa=msam10
 \font\ninemsa=msam9
 \font\sevenmsa=msam7
 \font\sixmsa=msam6
 \font\fourteenmsb=msbm10 scaled\magstep2
 \font\twelvemsb=msbm10 scaled\magstep1
 \font\tenmsb=msbm10
 \font\ninemsb=msbm9
 \font\sevenmsb=msbm7
 \font\sixmsb=msbm6
 \font\fourteeneu=eufm10 scaled\magstep2
 \font\twelveeu=eufm10 scaled\magstep1
 \font\teneu=eufm10
 \font\nineeu=eufm9
 
 \font\seveneu=eufm7
 \font\sixeu=eufm6
 \def\hexnumber@#1{\ifnum#1<10 \number#1\else
  \ifnum#1=10 A\else\ifnum#1=11 B\else\ifnum#1=12 C\else
  \ifnum#1=13 D\else\ifnum#1=14 E\else\ifnum#1=15 F\fi\fi\fi\fi\fi\fi\fi}
 \def\hexmsa{\hexnumber@\msafam}
 \def\hexmsb{\hexnumber@\msbfam} 
\fi
\newdimen\b@gheight             \b@gheight=12pt
\newcount\f@ntkey               \f@ntkey=0
\def\f@m{\afterassignment\samef@nt\f@ntkey=}
\def\samef@nt{\fam=\f@ntkey \the\textfont\f@ntkey\relax}
\def\rm{\f@m0 }
\def\mit{\f@m1 }
\def\cal{\f@m2 }
\def\it{\f@m\itfam}
\def\sl{\f@m\slfam}
\def\bf{\f@m\bffam}
\def\tt{\f@m\ttfam}
\def\caps{\f@m\cpfam}
\def\ssf{\f@m\ssfam}
\ifamsfonts
 \def\msa{\f@m\msafam}
 \def\msb{\f@m\msbfam} \let\bb=\msb
 \def\eu{\f@m\eufam}
\else
 \let \bb=\bf \let\eu=\bf
\fi
\def\fourteenpoint{\relax
    \textfont0=\fourteencp          \scriptfont0=\tenrm
      \scriptscriptfont0=\sevenrm
    \textfont1=\fourteeni           \scriptfont1=\teni
      \scriptscriptfont1=\seveni
    \textfont2=\fourteensy          \scriptfont2=\tensy
      \scriptscriptfont2=\sevensy
    \textfont3=\fourteenex          \scriptfont3=\twelveex
      \scriptscriptfont3=\tenex
    \textfont\itfam=\fourteenit     \scriptfont\itfam=\tenit
    \textfont\slfam=\fourteensl     \scriptfont\slfam=\tensl
      \scriptscriptfont\slfam=\sevensl
    \textfont\bffam=\fourteenbf     \scriptfont\bffam=\tenbf
      \scriptscriptfont\bffam=\sevenbf
    \textfont\ttfam=\fourteentt
    \textfont\cpfam=\fourteencp
    \textfont\ssfam=\fourteenss     \scriptfont\ssfam=\tenss
      \scriptscriptfont\ssfam=\sevenss
    \ifamsfonts
       \textfont\msafam=\fourteenmsa     \scriptfont\msafam=\tenmsa
         \scriptscriptfont\msafam=\sevenmsa
       \textfont\msbfam=\fourteenmsb     \scriptfont\msbfam=\tenmsb
         \scriptscriptfont\msbfam=\sevenmsb
       \textfont\eufam=\fourteeneu     \scriptfont\eufam=\teneu
         \scriptscriptfont\eufam=\seveneu \fi
    \samef@nt
    \b@gheight=14pt
    \setbox\strutbox=\hbox{\vrule height 0.85\b@gheight
                                depth 0.35\b@gheight width\z@ }}
\def\twelvepoint{\relax
    \textfont0=\twelverm          \scriptfont0=\ninerm
      \scriptscriptfont0=\sixrm
    \textfont1=\twelvei           \scriptfont1=\ninei
      \scriptscriptfont1=\sixi
    \textfont2=\twelvesy           \scriptfont2=\ninesy
      \scriptscriptfont2=\sixsy
    \textfont3=\twelveex          \scriptfont3=\ninex
      \scriptscriptfont3=\sixex
    \textfont\itfam=\twelveit    
    \textfont\slfam=\twelvesl    
      \scriptscriptfont\slfam=\sixsl
    \textfont\bffam=\twelvebf     \scriptfont\bffam=\ninebf
      \scriptscriptfont\bffam=\sixbf
    \textfont\ttfam=\twelvett
    \textfont\cpfam=\twelvecp
    \textfont\ssfam=\twelvess     \scriptfont\ssfam=\niness
      \scriptscriptfont\ssfam=\sixss
    \ifamsfonts
       \textfont\msafam=\twelvemsa     \scriptfont\msafam=\ninemsa
         \scriptscriptfont\msafam=\sixmsa
       \textfont\msbfam=\twelvemsb     \scriptfont\msbfam=\ninemsb
         \scriptscriptfont\msbfam=\sixmsb
       \textfont\eufam=\twelveeu     \scriptfont\eufam=\nineeu
         \scriptscriptfont\eufam=\sixeu \fi
    \samef@nt
    \b@gheight=12pt
    \setbox\strutbox=\hbox{\vrule height 0.85\b@gheight
                                depth 0.35\b@gheight width\z@ }}
\twelvepoint
%
%
\baselineskip = 15pt plus 0.2pt minus 0.1pt 
\lineskip = 1.5pt plus 0.1pt minus 0.1pt
\lineskiplimit = 1.5pt
\parskip = 6pt plus 2pt minus 1pt
\interlinepenalty=50
\interfootnotelinepenalty=5000
\predisplaypenalty=9000
\postdisplaypenalty=500
\hfuzz=1pt
\vfuzz=0.2pt
\dimen\footins=24 truecm 
\ifafour
 \hsize=16cm \vsize=22cm
\else
 \hsize=6.5in \vsize=9in
\fi
%
%
\skip\footins=\medskipamount
\newcount\fnotenumber
\def\clearfnotenumber{\fnotenumber=0} \clearfnotenumber
\def\fnote{\global\advance\fnotenumber by1 \generatefootsymbol
 \footnote{$^{\footsymbol}$}}
\def\fd@f#1 {\xdef\footsymbol{\mathchar"#1 }}
\def\generatefootsymbol{\iffrontpage\ifcase\fnotenumber
\or \fd@f 279 \or \fd@f 27A \or \fd@f 278 \or \fd@f 27B
\else  \fd@f 13F \fi
\else\xdef\footsymbol{\the\fnotenumber}\fi}
%
%
\newcount\secnumber \newcount\appnumber
\def\clearappnumber{\appnumber=64} \def\clearsecnumber{\secnumber=0}
\clearsecnumber \clearappnumber
\newif\ifs@c 
\newif\ifs@cd 
\s@cdtrue 
\def\unsectioned{\s@cdfalse\let\section=\subsection}
\newskip\sectionskip         \sectionskip=\medskipamount
\newskip\headskip            \headskip=8pt plus 3pt minus 3pt
\newdimen\sectionminspace    \sectionminspace=10pc
\def\Titlestyle#1{\par\begingroup \interlinepenalty=9999
     \leftskip=0.02\hsize plus 0.23\hsize minus 0.02\hsize
     \rightskip=\leftskip \parfillskip=0pt
     \advance\baselineskip by 0.5\baselineskip
     \hyphenpenalty=9000 \exhyphenpenalty=9000
     \tolerance=9999 \pretolerance=9000
     \spaceskip=0.333em \xspaceskip=0.5em
     \fourteenpoint
  \noindent #1\par\endgroup }
\def\titlestyle#1{\par\begingroup \interlinepenalty=9999
     \leftskip=0.02\hsize plus 0.23\hsize minus 0.02\hsize
     \rightskip=\leftskip \parfillskip=0pt
     \hyphenpenalty=9000 \exhyphenpenalty=9000
     \tolerance=9999 \pretolerance=9000
     \spaceskip=0.333em \xspaceskip=0.5em
     \fourteenpoint
   \noindent #1\par\endgroup }
\def\spacecheck#1{\dimen@=\pagegoal\advance\dimen@ by -\pagetotal
   \ifdim\dimen@<#1 \ifdim\dimen@>0pt \vfil\break \fi\fi}
\def\section#1{\cleareqnumber \s@ctrue \global\advance\secnumber by1
   \par \ifnum\the\lastpenalty=30000\else
   \penalty-200\vskip\sectionskip \spacecheck\sectionminspace\fi
   \noindent {\caps\enspace\S\the\secnumber\quad #1}\par
   \nobreak\vskip\headskip \penalty 30000 }
\def\undertext#1{\vtop{\hbox{#1}\kern 1pt \hrule}}
\def\subsection#1{\par
   \ifnum\the\lastpenalty=30000\else \penalty-100\smallskip
   \spacecheck\sectionminspace\fi
   \noindent\undertext{#1}\enspace \vadjust{\penalty5000}}

\def\appendix#1{\cleareqnumber \s@cfalse \global\advance\appnumber by1
   \par \ifnum\the\lastpenalty=30000\else
   \penalty-200\vskip\sectionskip \spacecheck\sectionminspace\fi
   \noindent {\caps\enspace Appendix \char\the\appnumber\quad #1}\par
   \nobreak\vskip\headskip \penalty 30000 }
\def\ack{\par\penalty-100\medskip \spacecheck\sectionminspace
   \line{\fourteencp\hfil ACKNOWLEDGEMENTS\hfil}%
\nobreak\vskip\headskip }
\def\refs{\begingroup \par\penalty-100\medskip \spacecheck\sectionminspace
   \line{\fourteencp\hfil REFERENCES\hfil}%
\nobreak\vskip\headskip \frenchspacing }
\def\endrefs{\par\endgroup}
%
%
\newif\iffrontpage \frontpagefalse
\headline={\hfil}
\footline={\iffrontpage\hfil\else \hss\twelverm
-- \folio\ --\hss \fi }
%
%
\newskip\frontpageskip \frontpageskip=12pt plus .5fil minus 2pt
\def\titlepage{\global\frontpagetrue\hrule height\z@ \relax
               \pubblock\relax }
\def\endtitlepage{\vfil\break\clearfnotenumber\frontpagefalse}
\def\title#1{\vskip\frontpageskip\Titlestyle{\caps #1}\vskip3\headskip}
\def\author#1{\vskip.5\frontpageskip\titlestyle{\caps #1}\nobreak}
\def\and{\par\kern 5pt \centerline{\sl and}}

\def\address#1{\par\kern 5pt\titlestyle{\it #1}}
\def\andaddress{\par\kern 5pt \centerline{\sl and} \address}

\def\abstract#1{\par\dimen@=\prevdepth \hrule height\z@ \prevdepth=\dimen@
   \vskip\frontpageskip\spacecheck\sectionminspace
   \centerline{\fourteencp ABSTRACT}\vskip\headskip
   {\noindent #1}}

\def\email#1{\fnote{\tentt e-mail: #1\hfill}}

%
%

%
\def\Bonn{\address{%
   Physikalisches Institut der Universit\"at Bonn\break
  Nu{\ss}allee 12, D--53115 Bonn, GERMANY}}
%

%

%
%
\newcount\refnumber \def\clearrefnumber{\refnumber=0}  \clearrefnumber
\newwrite\R@fs                              
\immediate\openout\R@fs=\jobname.refs 
\def\closerefs{\immediate\closeout\R@fs} 
\def\refsout{\closerefs\refs
\unlockat
\input\jobname.refs
\lockat
\endrefs}
\def\refitem#1{\item{{\bf #1}}}
\def\ifundefined#1{\expandafter\ifx\csname#1\endcsname\relax}
\def\[#1]{\ifundefined{#1R@FNO}%
\global\advance\refnumber by1%
\expandafter\xdef\csname#1R@FNO\endcsname{[\the\refnumber]}%
\immediate\write\R@fs{\noexpand\refitem{\csname#1R@FNO\endcsname}%
\noexpand\csname#1R@F\endcsname}\fi{\bf \csname#1R@FNO\endcsname}}
\def\refdef[#1]#2{\expandafter\gdef\csname#1R@F\endcsname{{#2}}}
%
%
\newcount\eqnumber \def\cleareqnumber{\eqnumber=0}
\newif\ifal@gn \al@gnfalse  
\def\veqnalign#1{\al@gntrue \vbox{\eqalignno{#1}} \al@gnfalse}
\def\eqnalign#1{\al@gntrue \eqalignno{#1} \al@gnfalse}
\def\(#1){\relax%
\ifundefined{#1@Q}
 \global\advance\eqnumber by1
 \ifs@cd
  \ifs@c
   \expandafter\xdef\csname#1@Q\endcsname{{%
\noexpand\rm(\the\secnumber .\the\eqnumber)}}
  \else
   \expandafter\xdef\csname#1@Q\endcsname{{%
\noexpand\rm(\char\the\appnumber .\the\eqnumber)}}
  \fi
 \else
  \expandafter\xdef\csname#1@Q\endcsname{{\noexpand\rm(\the\eqnumber)}}
 \fi
 \ifal@gn
    & \csname#1@Q\endcsname
 \else
    \eqno \csname#1@Q\endcsname
 \fi
\else%
\csname#1@Q\endcsname\fi\global\let\@Q=\relax}
%
%
\newif\ifm@thstyle \m@thstylefalse
\def\mathstyle{\m@thstyletrue}
\def\proclaim#1#2\par{\smallbreak\begingroup
\advance\baselineskip by -0.25\baselineskip%
\advance\belowdisplayskip by -0.35\belowdisplayskip%
\advance\abovedisplayskip by -0.35\abovedisplayskip%
    \noindent{\caps#1.\enspace}{#2}\par\endgroup%
\smallbreak}
\def\m@kem@th<#1>#2#3{%
\ifm@thstyle \global\advance\eqnumber by1
 \ifs@cd
  \ifs@c
   \expandafter\xdef\csname#1\endcsname{{%
\noexpand #2\ \the\secnumber .\the\eqnumber}}
  \else
   \expandafter\xdef\csname#1\endcsname{{%
\noexpand #2\ \char\the\appnumber .\the\eqnumber}}
  \fi
 \else
  \expandafter\xdef\csname#1\endcsname{{\noexpand #2\ \the\eqnumber}}
 \fi
 \proclaim{\csname#1\endcsname}{#3}
\else
 \proclaim{#2}{#3}
\fi}
\def\Thm<#1>#2{\m@kem@th<#1M@TH>{Theorem}{\sl#2}}
\def\Prop<#1>#2{\m@kem@th<#1M@TH>{Proposition}{\sl#2}}
\def\Def<#1>#2{\m@kem@th<#1M@TH>{Definition}{\rm#2}}
\def\Lem<#1>#2{\m@kem@th<#1M@TH>{Lemma}{\sl#2}}
\def\Cor<#1>#2{\m@kem@th<#1M@TH>{Corollary}{\sl#2}}
\def\Conj<#1>#2{\m@kem@th<#1M@TH>{Conjecture}{\sl#2}}
\def\Rmk<#1>#2{\m@kem@th<#1M@TH>{Remark}{\rm#2}}
\def\Exm<#1>#2{\m@kem@th<#1M@TH>{Example}{\rm#2}}
\def\Qry<#1>#2{\m@kem@th<#1M@TH>{Query}{\it#2}}
\def\Proof{\noindent{\caps Proof:}\enspace}
\def\<#1>{\csname#1M@TH\endcsname}
%
%
\def\ref#1{{\bf [#1]}}
\def\ie{{\it i.e.\/}}
\def\eg{{\it e.g.\/}}
\def\nl{\hfil\break}
%
%
\def\qed{\vrule width 0.7em height 0.6em depth 0.2em}
\def\QED{\enspace\qed}
\def\lapprox{\hbox{\lower3pt\hbox{$\buildrel<\over\sim$}}}
\def\gapprox{\hbox{\lower3pt\hbox{$\buildrel<\over\sim$}}}
\def\quotient#1#2{#1/\lower0pt\hbox{${#2}$}}
\def\fr#1/#2{\mathord{\hbox{${#1}\over{#2}$}}}
\ifamsfonts
 \mathchardef\empty="0\hexmsb3F 
 \mathchardef\lsemidir="2\hexmsb6E 
 \mathchardef\rsemidir="2\hexmsb6F 
\else
 \let\empty=\emptyset
 \def\lsemidir{\mathbin{\hbox{\hskip2pt\vrule height 5.7pt depth -.3pt
    width .25pt\hskip-2pt$\times$}}}
 \def\rsemidir{\mathbin{\hbox{$\times$\hskip-2pt\vrule height 5.7pt
    depth -.3pt width .25pt\hskip2pt}}}
\fi
%
\def\to{\rightarrow}
\def\tto{\longrightarrow}
\def\lra{\leftrightarrow}
\def\mapright#1{\smash{
    \mathop{\tto}\limits^{#1}}}

\def\mapdown#1{\Big\downarrow
  \rlap{$\vcenter{\hbox{$\scriptstyle#1$}}$}}

\def\commdiag#1{
\def\normalbaselines{\baselineskip20pt \lineskip3pt \lineskiplimit3pt }
\matrix{#1}} 
%
\def\integ{\mathord{\bb Z}} 
%
\def\ker{\mathop{\rm ker}}
%
\def\underrightarrow#1{\vtop{\ialign{##\crcr
      $\hfil\displaystyle{#1}\hfil$\crcr
      \noalign{\kern-\p@\nointerlineskip}
      \rightarrowfill\crcr}}} 
\def\underleftarrow#1{\vtop{\ialign{##\crcr
      $\hfil\displaystyle{#1}\hfil$\crcr
      \noalign{\kern-\p@\nointerlineskip}
      \leftarrowfill\crcr}}}  

%
%
\def\pder#1#2{{{\partial #1}\over{\partial #2}}}
\def\der#1#2{{{d #1}\over {d #2}}}
%
%

\def\CMP#1#2#3{{\sl Comm. Math. Phys.} {\bf #1} (#2) #3}

\def\PLB#1#2#3{{\sl Phys. Lett.} {\bf #1B} (#2) #3}

\def\FAaIA#1#2#3{{\sl Functional Analysis and Its Application} {\bf #1} (#2)
#3}

\def\Invm#1#2#3{{\sl Invent. math.} {\bf #1} (#2) #3}
\def\LMP#1#2#3{{\sl Letters in Math. Phys.} {\bf #1} (#2) #3}

\def\RMaP#1#2#3{{\sl Reports on Math. Phys.} {\bf #1} (#2) #3}

\def\MPLA#1#2#3{{\sl Mod. Phys. Lett.} {\bf A#1} (#2) #3}

\def\JETPL#1#2#3{{\sl  Sov. Phys. JETP Lett.} {\bf #1} (#2) #3}
\def\JDG#1#2#3{{\sl J. Diff. Geometry} {\bf #1} (#2) #3}

\lockat
%
%
\def\ord{\mathop{\rm ord}}

\let\Z=\integ

\def\SPDO{\mathord{{\ssf S}\Psi{\ssf DO}}}
\def\SDOP{\mathord{\ssf SDOP}}
\def\sbin[#1,#2]{\left[{#1\atop #2}\right]}
\let\d=\partial

\def\gM{{\eu M}}
\def\Str{\mathop{\rm Str}}
\def\sres{\mathop{\rm sres}}
\def\dlbra{[\![}
\def\drbra{]\!]}
\def\SKP2{$\hbox{SKP}_2$}
\def\DOP{\mathord{\ssf DOP}}
\def\SW{\mathord{\ssf SW}}
\def\W{\mathord{\ssf W}}
\ifamsfonts
 \mathchardef\whitesq="0\hexmsa03
 \def\QED{\hfill$\whitesq$}
\else
 \def\QED{\hfill\qed}
\fi
\def\fr#1/#2{\mathord{\hbox{${#1}\over{#2}$}}}
\def\half{\fr1/2}
\refdef[Radul]{A.~O.~Radul, \JETPL{50}{1989}{371},
\FAaIA{25}{1991}{25};\nl
I.~Vaysburd \& A.~Radul, \PLB{274}{1992}{317}.}
\refdef[AK]{S.~Aoyama \& Y.~Kodama, \PLB{278}{1992}{56}.}
\refdef[DB]{L.~A.~Dickey,  {\sl Soliton equations and Hamiltonian
systems}, Advanced Series in Mathematical Physics Vol.12,  World
Scientific Publ.~Co..}
\refdef[OS]{A.Yu. Orlov \& E.I. Shulman, \LMP{12}{1986}{171}.}
\refdef[qRadul]{J.M.Figueroa-O'Farrill \& E. Ramos, {\tt
hep-th/9211036}, \LMP{27}{1993}{223}.}
\refdef[MRSKP]{Yu. I. Manin \& A. O. Radul, \CMP{98}{1985}{65}.}
\refdef[MulaseMR]{M. Mulase, \Invm{92}{1988}{1}.}
\refdef[MulaseJ]{M. Mulase, \JDG{34}{1991}{651}.}
\refdef[Rabin]{J. Rabin, \CMP{137}{1991}{533}.}
\refdef[Das]{A.~Das, E.~Sezgin, \& S.~J.~Sin, {\tt hep-th/9111054},
\PLB{277}{1992}{435}.}
\refdef[eSKdV]{J.M. Figueroa-O'Farrill, J. Mas \& E. Ramos,
\RMaP{3}{1991}{479}.}
\refdef[Adler]{M.~Adler, \Invm{50}{1979}{403}.}
\refdef[DickeyAS]{L.~A.~Dickey, {\sl Additional Symmetries of KP,
Grassmannian, and the String Equation}, {\tt hep-th/9204092}; {\sl
Part II} {\tt hep-th/9210155}.}
\refdef[JECMP]{J.M. Figueroa-O'Farrill \& E. Ramos,
\CMP{145}{1992}{43}.}
\refdef[Beckers]{K. Becker \& M. Becker, {\tt hep-th/9301017},
\MPLA{8}{1993}{1205}.}
\refdef[SKdVB]{J.M. Figueroa-O'Farrill \& S. Stanciu, {\tt
hep-th/9302057}, \PLB{316}{1993}{282}.}
\refdef[SBouss]{J.M. Figueroa-O'Farrill \& S. Stanciu, {\tt
hep-th/9303168}, \MPLA{8}{1993}{2125}.}
\refdef[MMReview]{P. Di Francesco, P. Ginsparg, \& J. Zinn-Justin,
{\sl $2D$ Gravity and Random Matrices}, {\tt hep-th/9306153}, to appear in
{\it Physics Reports}.}
\refdef[SWinfty]{E. Bergshoeff, C.N. Pope, L.J. Romans, E. Sezgin, \&
X. Shen, \PLB{245}{1990}{447}.}

\overfullrule=0pt
\mathstyle
%
\def\pubblock{ \line{\hfil\rm BONN--HE--93--31}
               \line{\hfil\tt hep-th/9309058}
               \line{\hfil\rm September 1993}
               \line{\hfil\rm March 1994 (revised)}}
\titlepage
\title{Additional Symmetries of Supersymmetric KP Hierarchies}
\author{Sonia Stanciu\email{sonia@avzw02.physik.uni-bonn.de}}
\Bonn
\abstract{We investigate the additional symmetries of several
supersymmetric KP hierarchies: the SKP hierarchy of Manin and Radul,
the $\hbox{SKP}_2$ hierarchy, and the Jacobian SKP hierarchy.  In all
three cases we find that the algebra of symmetries is isomorphic to
the algebra of superdifferential operators, or equivalently
$\SW_{1+\infty}$.  These results seem to suggest that despite their
realization depending on the dynamics, the additional symmetries are
kinematical in nature.}
\endtitlepage

\section{Introduction}

The aim of this paper is to study the additional symmetries of
the following supersymmetric extensions of the KP hierarchy: the SKP
hierarchy of Manin and Radul \[MRSKP], the $\hbox{SKP}_2$ hierarchy
\[eSKdV], and the Jacobian SKP hierarchy of Mulase \[MulaseJ] and
Rabin \[Rabin].  The additional symmetries of the KP hierarchy were
first studied in \[OS] (although see \[DickeyAS]) and their algebraic
structure has been recently identified with the algebra $\DOP$ of
differential operators \[AK]\[qRadul].  The additional symmetries of
the supersymmetric KP hierarchy defined by Manin and Radul have been
previously studied in \[Das].

Let us first briefly review what is known about the additional
symmetries of the KP hierarchy.   The KP hierarchy can be thought of
as a dynamical system defined on a space whose functions are given by
a subring of the ring $k[[x]]$ of formal power series in the variable
$x$ with coefficients in the field $k$.  It is defined as the
universal family of isospectral deformations of a pseudodifferential
operator $L=\d +\sum_{i\geq 1}u_i\d^{1-i}$.  The evolution of $L$ is
specified by a commuting family of flows $\d_i$ in terms of which
$$
\d_i L = -[L_-^i,L] = [L_+^i,L]~.\()
$$
If one restricts oneself to operators satisfying $u_1=0$, then one can
lift the KP flows to the Volterra group $G$.  The Volterra group acts
naturally via dressing transformations $L\mapsto \phi^{-1}L\phi$,
where $\phi= 1+\sum_{i\geq 1}v_i\d^{-i} \in G$ is the dressing
operator.  In terms of the dressing operator, the flows of the KP
hierarchy are given by
$$
\d_i\phi = -\left(\phi\d^i\phi^{-1}\right)_-\phi~.\(KPflows)
$$

One can write these flows in a different way by using an analogue of
the Radul map \[Radul].  Motivated originally by attempts to
understand the geometric meaning of $\W$-symmetry, Radul introduced a
homomorphism
$$
W : \DOP \to T_L\gM~,\(Radul)
$$
between the Lie algebra of differential operators and the Lie algebra
of vector fields on the space of Lax operators by associating, to
every differential operator $E$, the tangent vector
$$
W(E)=(LEL^{-1})_- L~\(NotRadul)
$$
at the point $L$ in the space $\gM$ of Lax operators.  The similarity
between the expression for the Radul map \(Radul) and the one of the KP
flows \(KPflows) suggests us to define a map
$$
W'(E)=(\phi E\phi^{-1})_-\phi~,\(w)
$$
from $\DOP$ to the Lie algebra ${\cal R}_-$ of the Volterra group.
The KP flows become now $\d_n\phi=-W'(\d^n)=-\d_{W'(\d^n)}\phi$,
where $\d_{W'(\d^n)}$ is then a flow on the Volterra group. The
map \(w) now translates the trivial fact $[\d^n,\d^m]=0$ into the
commutativity of the flows $[\d_n,\d_m]=0$.  This allows us to
represent the flows in terms of an infinite set of times,
$\d_i=\pder{}{t_i}$, with $i=1,2,\ldots$.  One interpretation of this
feature is that every flow possesses an infinite number of symmetries
given by the other flows.  This interpretation begs the question
whether these are all or, if on the contrary, there exist additional
symmetries.  Remarkably enough, it turns out that one can construct a
larger family of times-dependent flows which contains as a subset the
original KP flows and commute with them.  This new family of flows
satisfies a nonabelian algebra with respect to which the KP hierarchy
forms its center.  Thus we adopt here the following definition.

\Def<addsymdef>{By (additional) symmetries of an integrable hierarchy
of flows, we mean its centralizer in the algebra of times-dependent
vector fields.}

The fact that these symmetries contain the original hierarchy,
although largely taken for granted, is only true provided the flows of
the hierarchy themselves satisfy an abelian algebra; and it is to
these cases that the word `additional' can be applied.  We will see in
fact that this is not generally the case for supersymmetric
hierarchies.

Along with \<addsymdef>, it is in practice convenient to have a
`working definition' that is more suitable for computation.  Our
working definition is motivated by the following fact.  The flows
$\d_{W'(\Gamma)}$ generated via \(w) by differential operators
$\Gamma$ satisfying
$$
[\d_i - \d^i, \Gamma] = 0\(addsym)
$$
commute with the KP flows.  Indeed following \[qRadul] we have that
$$
\eqnalign{
[\d_{W'(\Gamma)},\d_i]&=-[\d_{W'(\Gamma)},\d_{W'(\d^i)}]\cr
                      &=-\d_{W'([\Gamma,\d^i]_{\phi})}~,\()\cr}
$$
where $[\Gamma,\d^i]_{\phi}$ is a modified Lie bracket (see Sect.3)
given by
$$
\veqnalign{
[\Gamma,\d^i]_{\phi}&=\d_{W'(\Gamma)}\d^i-\d_{W'(\d^i)}\Gamma+
                                                    [\Gamma,\d^i]\cr
                    &=[\d_i,\Gamma]+[\Gamma,\d^i]\cr
                    &=[\d_i - \d^i,\Gamma]\cr
                    &=0~.\()\cr}
$$
We therefore call `additional symmetries' the flows generated by
operators $\Gamma$ subject to \(addsym).  It is conceivable that
\(addsym) is also a necessary condition---that is, that all additional
symmetries arise in this fashion; but we shall not attempt to prove it
here.

This means that looking for the additional symmetries comes down to
trying to find solutions for \(addsym). An obvious solution to this
equation is simply $\Gamma=\d$, which, introduced in \(addsym) and after
applying a dressing transformation, gives precisely the KP flows
\(KPflows).  This agrees with the fact that the KP flows commute with
each other.

A more interesting solution can be obtained if we allow for an
explicit dependence on the time parameters of the hierarchy; that is,
if we extend our ring of functions by the infinite set of independent
variables $\{t_1,t_2,\ldots\}$, in which case we have to extend the
derivative operator $\d$ as a derivation in this new ring.

A priori, since $x$ and all the $t_i$ are independent variables of our
infinite set of partial differential equations, we can automatically
conclude that $\d$ has to be extended trivially to the new ring.
Nevertheless in the case KP, since the first flow (for dressable $L$)
reads $\d_1 L=[L_+,L]=[\d,L]$ and therefore gives $\d=\d_1$, one can
identify $x$ with $t_1$. One can then define (see, for instance,
\[DB]) a formal infinite-order differential operator
$$
\Gamma = \sum_{j\geq 1}jt_j\d^{j-1}~,\()
$$
which satisfies \(addsym), and from it a two-parameter family of flows
$$
\d_{m,k}\phi = \left(\phi\Gamma^k\d^m\phi^{-1}\right)_-\phi~,\()
$$
that satisfy $[\d_{m,k},\d_n]=0$.  Notice that for $k=0$ and $m>0$
they agree with the KP flows.  Moreover since $[\d,\Gamma]=1$ it
follows that the Lie algebra generated by $\Gamma^k\d^m$, $k\geq 0$
and $m\in\Z$ is isomorphic (as a Lie algebra) to $\W_{1+\infty}$ and
hence the algebra of additional symmetries is nothing but
$\W_{\infty}$ (see \[AK] and \[qRadul]).

One can alternatively write the two-parameter family of flows in a Lax
form
$$
\d_{m,k}L = -[(M^k L^m)_-,L]~,
$$
where $M=\phi\Gamma\phi^{-1}$ is the dressed version of $\Gamma$.  In
this form, the additional symmetries prove to be a useful tool in the
solution of the (multi-)matrix models for $2d$ quantum gravity.
Indeed, the partition function of the $(n-1)$-matrix model can be
identified with the $\tau$-function of the $n$-KdV hierarchy.  The
$\tau$-function is subject to an infinite set of $W$-constraints which
follow from the string equation $[P,Q]=1$, where $Q$ is the $n$-th
power of the Lax operator for the KP hierarchy and $P={1\over
n}(ML^{-n+1})_+$ is directly related to the generator $M$ of the
additional symmetries \[MMReview].

In this paper we analyze the additional symmetries of several
supersymmetric extensions of the KP hierarchy along these same lines.
In order to do that we shall start in the second section with a brief
review of a few basic facts about the supersymmetric formalism of
pseudodifferential operators in order to introduce the basic objects
we shall work with and to fix the notation. (For a more detailed
account on this topic we send the reader to \[MRSKP], \[MulaseMR],
\[MulaseJ], and \[eSKdV].)

As we have seen in the case of the KP hierarchy, a very useful tool in
the study of the additional symmetries is the map \(NotRadul). We
shall therefore need to define its supersymmetric version.
We do that in the third section where we also prove that, analogously
to the nonsupersymmetric case, it is a Lie algebra homomorphism.

In the fourth section we will consider the additional symmetries of
the SKP hierarchy introduced by Manin and Radul (MRSKP).  One of the
distinctive features of this supersymmetric hierarchy is the fact that
its algebra of flows is nonabelian and therefore not all the flows of
the hierarchy are symmetries as well.  Imposing an analogous condition
to \(addsym) we shall find the algebra of the additional symmetries to
be isomorphic with the algebra of superdifferential operators $\SDOP$.
The additional symmetries of this particular hierarchy have been
studied also in \[Das] and we find agreement with their results.

The even order SKP hierarchy will be discussed in the fifth section.
An important feature of this hierarchy is the fact that it has only
even Lax flows which can be represented in terms of an infinite set of
even times.  This has as a consequence the fact that not only the
generator $Q$ of supertranslations but also the odd derivation $D$
generate symmetries of the hierarchy.  As a result, the algebra of
additional symmetries will be again seen to be the algebra $\SDOP$ of
superdifferential operators.

Finally, in the sixth section we shall consider the Jacobian SKP
hierarchy of Mulase and Rabin (JSKP), and we shall find its
additional symmetries.  In contrast to MRSKP, the flows of JSKP do
commute with each other and the additional flows will therefore
contain the JSKP flows as well.  Moreover, we shall find that the
symmetries form a Lie algebra isomorphic to the Lie algebra of
superdifferential operators, which is in turn isomorphic---as a Lie
algebra---with $\SW_{1+\infty}$; the additional flows corresponding to
a subalgebra containing $\SW_\infty$.

The results seem to suggest that the additional symmetries considered
so far are to a large extent kinematical in nature, although their
explicit realization does depend on the dynamics.

\section{Supersymmetric formalism}

Let $k$ be an arbitrary field of characteristic zero. We define our
function space to be a $\Z_2$-graded ring over $k$, $R=R_0\oplus R_1$,
endowed with an odd derivation $D$. Then $D^2=\d$ is an even
derivation. For the present purposes it suffices to restrict ourselves
to the case in which $R$ is given by
$$
R = K\oplus K\theta~,\()
$$
where $K$ is a subring of $k[[x]]$, the ring of formal power series in
an even variable $x$, and $\theta$ is an odd variable satisfying
$x\theta=\theta x$ and $\theta^2=0$. This gives rise to a structure of
supercommutative algebra in the ring $R$. The $\Z_2$-grading is
defined by putting $|x|=0$ and $|\theta|=1$ so that any element of
$R_0$ (respectively $R_1$) is homogeneous of degree $0$ (respectively
$1$). The odd derivation operator is given by
$D=\d_{\theta}+\theta\d$ and satisfies the supersymmetric analog of
the Leibniz rule
$$
D(ab) = D(a)b + (-)^{|a|}aD(b)~,\()
$$
where $a$ is a homogeneous element of $R$ of $\Z_2$-degree $|a|$ and
$|D|=1$. We further define the ring of supersymmetric
pseudodifferential operators ($\SPDO$) with coefficients in $R$
$$
{\cal R}\equiv R((D^{-1})) = \left\{\left. P=\sum_{i\gg -\infty}a_i
D^{-i} \right| a_i\in R \right\}~.\()
$$
The ring of $\SPDO$'s can be given the structure of a superalgebra
using the generalized Leibniz rule
$$
D^k a = \sum_{i=0}^{\infty}\sbin[k,k-i] (-1)^{|a|(k-i)}a^{[i]}
D^{k-i}~,\()
$$
where $a$ is a homogeneous element of $R$ and $\sbin[k,k-i]$ are the
so-called superbinomial coefficients given by
$$
\sbin[k,k-i] = \cases{0&for $i<0$ or $(k,i)\equiv (0,1)\bmod{2}$;\cr
{\left({{\left[\fr k/2\right]}\atop{\left[\fr k-i/2\right]}}\right)}&for
$i\geq 0$ and $(k,i)\not\equiv(0,1)\bmod{2}$.\cr}
$$
Since the $\Z_2$-grading gets induced here we have that ${\cal
R}={\cal R}_0\oplus{\cal R}_1$ where
$$
\eqnalign{
{\cal R}_0 &= R_0((D^{-1})) =\left\{\left. \sum_{i\gg-\infty}a_i
D^{-i} \right| |a_{2i}|=0\ ,|a_{2i+1}|=1 \right\}~,\()\cr
\noalign{\hbox{and}}
{\cal R}_1 &= R_1((D^{-1})) =\left\{\left. \sum_{i\gg-\infty}a_i
D^{-i} \right| |a_{2i}|=1\ ,|a_{2i+1}|=0 \right\}~,\()\cr}
$$
and we have thus defined the notion of an even (respectively odd)
$\SPDO$.

Let us remark the following fact:
$$
R((D^{-1})) = R((\d^{-1}))\oplus R((\d^{-1}))\d_{\theta}~.\()
$$
Indeed, on the one hand we clearly have $R((\d^{-1}))\oplus
R((\d^{-1}))\d_{\theta}\subset R((D^{-1}))$ since $D^2=\d$ and
$\d_{\theta}=D-\theta D^2$. On the other hand any $\SPDO$ can be
written in the following manner:
$$
\eqnalign{
\sum_i a_i D^i &=\sum_i a_{2i}D^{2i} + \sum_i a_{2i+1}D^{2i+1}\cr
               &=\sum_i a_{2i}\d^i + \sum_i
                            a_{2i+1}\d^i(\d_{\theta}+\theta\d)\cr
               &=\sum_i\left(a_{2i}+a_{2i+1}\theta\right)\d^i +
                 \sum_i a_{2i-1}\d^i\d_{\theta}~,\()\cr}
$$
so that we have also $R((D^{-1}))\subset R((\d^{-1}))\oplus
R((\d^{-1}))\d_{\theta}$.

In general it is important to distinguish in the ring of $\SPDO$'s the
subring of supersymmetric differential operators $\SDOP$
$$
{\cal R}_+\equiv R[D] = \left\{\left.\sum_{0\leq i \ll \infty} a_i
                      D^i\right| a_i\in R\right\}~,\()
$$
with respect to which we have the splitting
$$
{\cal R} = {\cal R}_+ \oplus {\cal R}_-~,\()
$$
where
$$
{\cal R}_- \equiv D^{-1}R[[D^{-1}]] = \left\{\left.\sum_{i=1}^{\infty}a_i
D^{-i} \right| a_i\in R\right\}\()
$$
denotes the integral $\SPDO$'s.  If $P\in{\cal R}$ is any $\SPDO$ we
shall denote by $P_\pm$ its projection onto ${\cal R}_\pm$ along
${\cal R}_\mp$.

The ring ${\cal R}$ of $\SPDO$'s can be made into a filtered
associative $k$-algebra if we define
$$
{\cal R}^n= \left\{\left. P=\sum_i a_i D^i \in {\cal R}\right| \ord P
            \leq n\right\}~,\()
$$
where we say that $\ord P=n$ if $a_i=0$ for all $i>n$ and $a_n\not=
0$. We have then
$$
{\cal R}^n\subset {\cal R}^{n+1}\ \ {\rm and}\ \  {\cal R} =
\cup_{n\in\Z} {\cal R}^n \()
$$
and under the multiplication ${\cal R}^p\times {\cal R}^q \to {\cal
R}^{p+q}$ . Moreover defining a bracket
$$
[~~]: {\cal R}^p\times {\cal R}^q\to {\cal R}^{p+q}\()
$$
via the graded commutator $[AB]=AB-(-)^{|A||B|}BA$, ${\cal R}$ becomes
a Lie superalgebra.

Let us now consider the supersymmetric trace map $\Str =
\int\circ\sres$ where the supersymmetric noncommutative residue is
given by $\sres\left(\sum_i a_i D^i\right)=a_{-1}$ and the notion of
integration can be defined abstractly as the canonical projection
$\int :R\to R/DR$ which simply means dropping the derivatives.
One can alternatively consider the integral defined by the Berezinian,
where for any homogeneous differential polynomial of $U=u+\theta v$,
$f(U)=a(u,v)+\theta b(u,v)$, $\int_B f(U)=\int b(u,v)$. The only
difference between the two definitions consists in the fact that the
abstract integration defined directly as a canonical projection is an
even operation whereas the Berezinian has a $\Z_2$-degree of one.

The $\Str$ functional allows us now to define a dual pairing in ${\cal
R}$ by
$$
<A,B>\equiv \Str AB~,\(pairing)
$$
under which $R[D]$ and $D^{-1}R[[D^{-1}]]$ are maximally isotropic
spaces nondegenerately paired with each other.

One of the central objects in our formalism is the space of
supersymmetric Lax operators of degree $n$, defined by
$$
\gM_n\equiv \left\{\left. L=D^n+\sum_{i=1}^{\infty}U_i D^{n-i} \right|
U_i\in R, |U_i|\equiv n+i \bmod 2\right\}~.\()
$$
(We shall drop the subscript $n$ whenever no confusion can arise.)
Given any $L\in \gM \subset {\cal R}$ one can define $R_L$ the
differential subring generated by the $U_i$'s which will obviously
induce the corresponding subrings $R_L [D]\subset R[D]$ and
$R_L[D^{-1}]\subset R[D^{-1}]$. $\gM$ is an infinite-dimensional
affine space modeled on the vector space ${\cal R}^{n-1}$ of $\SPDO$'s
of order $n-1$. Its tangent space $T_L\gM$ at the point $L$ is
thus isomorphic to ${\cal R}^{n-1}$ itself, namely
$$
T_L\gM\equiv\left\{\left. A=\sum_{i=1}^{\infty}A_i D^{n-i} \right|
A_i\in R, |A_i|\equiv |A|+i \bmod 2\right\}~.\()
$$
Then to every tangent vector $A\in T_L\gM$ one can associate a vector
field $D_A$ whose action on any $f\in R_L$ is defined by
$$
\eqnalign{
D_A f&\equiv\left.\der{}{\epsilon}f\left(U_i+\epsilon
A_i\right)\right|_{\epsilon=0}\cr
          &=\sum_{i=1}^{\infty}\sum_{k=0}^{\infty}(-)^{(|A|+n)k}A_i^{[k]}
            \pder{f}{U_i^{[k]}}~,\()\cr}
$$
where we do not impose a priori that $\epsilon$ be even, \ie, $|L|=|A|$
is not necessarily true.

\Lem<evolder>{$D_A D = (-)^{(|A|+n)} D D_A$.}

The proof is a straightforward computation and it is left as an
exercise.

Notice that $D_A :R_L\to R_L$ induces a map---also denoted $D_A$ with
some abuse of notation---$D_A :R_L/DR_L\to R_L/DR_L$ given by
$D_A\int f = \int D_A f$.  But in this case since $R_L/DR_L$ is no
longer a superalgebra, $D_A$ is no longer a derivation. This
nevertheless will not affect the formalism.

In order to define the cotangent space of $\gM$ at $L$ it suffices to
notice that the tangent space $T_L\gM$ is nondegenerately paired (via
the pairing defined in \(pairing)) with the quotient space ${\cal
R}/D^{-n}{\cal R}_-$ and therefore we have
$$
T_L^*\gM\cong {\cal R}/D^{-n}{\cal R}_-~.\()
$$

\section{The supersymmetric Radul map}

In this section we shall introduce a supersymmetric generalization of
the Radul map and we shall see that it defines a Lie algebra
homomorphism between the space of $\SDOP$'s and $T_L\gM$. In order to
do this we have first of all to define a Lie (super)algebra structure
on $T_L\gM$. Of course, since the elements of $T_L\gM$ are in
particular $\SPDO$'s of order at most $n-1$ one always has the obvious
bracket given by the graded commutator. Still this is not the one that
will allow us to exhibit the supersymmetric Radul map as a Lie algebra
homomorphism. Instead let us consider the natural Lie bracket on
vector fields on $\gM$, namely
$$
\left[ D_A, D_B\right] =  D_A D_B -(-)^{| D_A|| D_B|} D_B D_A~.\()
$$
This will induce in $T_L\gM$ a bracket $\dlbra~\drbra$ by
$$
\left[ D_A, D_B\right] =  D_{\dlbra A,B\drbra}~,\()
$$
whose explicit form we shall obtain now.

\Lem<>{The Lie bracket $\dlbra~~\drbra$ in $T_L\gM$ is given by
$$
\dlbra A,B \drbra =  D_A B-(-)^{| D_A| |D_B|} D_B A~,\(lemma)
$$
for any two $\SPDO$'s $A$ and $B$ in $T_L\gM$.}

\Proof Consider $f$ an arbitrary function in $R_L$. Then
$$
\left[ D_A, D_B\right]f =\left( D_A D_B -(-)^{| D_A|| D_B|} D_B
D_A\right)f~,\()
$$
which using the fact that $ D_A f^{[k]}=(-)^{k| D_A|}\left( D_A
f\right)^{[k]}$ (which follows by repeated application of \<evolder>)
becomes
$$
\eqnalign{
\left[ D_A, D_B\right]f ={}& \sum_{i,j=1}^{\infty}
\sum_{k,l=0}^{\infty}\left((-)^{k| D_B|+(l+k)| D_A|}
A_j^{[l]}\pder{B_i}{U_j^{[k]}}\right.\cr
&\left.\qquad -(-)^{| D_A|| D_B| + k| D_A| +
(l+k)| D_B|} B_j^{[l]}
\pder{A_i}{U_j^{[l]}}\right)^{[k]}\pder{f}{U_i^{[k]}}\cr
={}&\sum_{i=1}^{\infty}\sum_{k=0}^{\infty}(-)^{k(| D_A|+|
D_B|)} \dlbra A,B\drbra_i^{[k]}\pder{f}{U_i^{[k]}}~,\()\cr}
$$
where
$$
\dlbra A,B\drbra_i = \sum_{j=1}^{\infty}
\sum_{l=0}^{\infty}(-)^{l| D_A|}A_j^{[l]}\pder{B_i}{U_j^{[k]}}
-(-)^{| D_A|| D_B| + l| D_B|} B_j^{[l]}
\pder{A_i}{U_j^{[l]}}~,\()
$$
and we get indeed \(lemma). \QED

It is well known \[DB] that one can pull the Lie bracket
$\dlbra~~\drbra$ on $T_L\gM$ back to $T_L^*\gM$ via the Adler map
$J(X)=(LX)_+ L - L(XL)_+$. In other words one can define a bracket
$[~~]^*_L$ on $T_L^*\gM$ such that
$$
\dlbra J(X),J(Y)\drbra = J([X,Y]_L^*)~.\()
$$
Computing $[X,Y]^*_L$ one finds \[JECMP]
$$
\eqnalign{
[X,Y]_L^* = &(-)^{n(n+|X|)} D_{J(X)}Y + X(LY)_-\cr
            &- ((XL)_+ Y)_- -(-)^{(n+|X|)(n+|Y|)} (X \lra Y)~.\()\cr}
$$
This already tells us that the Adler map is a Lie (super)algebra
homomorphism mapping the cotangent space to the tangent space of $\gM$
at $L$, each of them being considered with the corresponding Lie
algebra structure.

Now we can finally define the supersymmetric analog of the Radul map
$$
W : \SDOP \to T_L\gM\()
$$
sending any $E\in\SDOP$ to the tangent vector $W(E)$ defined by
$$
W(E) \equiv LE - (LEL^{-1})_+ L = (LEL^{-1})_- L~.\()
$$

\Thm<SRadul>{The supersymmetric Radul map is a Lie
(super)algebra homomorphism, \ie,
$$
\dlbra W(E),W(F)\drbra = W([E,F]_L)~,\()
$$
where $[E,F]_L$ is the modified Lie bracket on $\SDOP$ given by
$$
[E,F]_L = [E,F] + (-)^{n|E|} D_{W(E)}F -
(-)^{|E||F|+n|F|} D_{W(F)}E~. \()
$$
}

\Proof
By direct computation in the right hand side we have
$$
\eqnalign{
\dlbra W(E),W(F)\drbra \equiv{}&  D_{W(E)}W(F) -
(-)^{| D_{W(E)}|| D_{W(F)|}} D_{W(F)}W(E)\cr
={}&  D_{W(E)}(LFL^{-1})_- L -(-)^{|
D_{W(E)}|| D_{W(F)}|}(E\lra F)\cr
={}& (W(E)FL^{-1})_- L + (-)^{n| D_{W(E)}|}(L{
D}_{W(E)}FL^{-1})_- L \cr
& -(-)^{|F|| D_{W(E)}|}(LFL^{-1}W(E)L^{-1})_- L\cr
& + (-)^{|F|| D_{W(E)}|}(LFL^{-1})_- W(E) - (-)^{|
D_{W(E)}|| D_{W(F)}|} (E\lra F)\cr
={}& ((LEL^{-1})_- LFL^{-1})_- L + (-)^{|E||F|}((LFL^{-1})_-
LEL^{-1})_- L \cr
& - (-)^{|E||F|}(LFEL^{-1})_- L + (-)^{n|E|}(L D_{W(E)}
  FL^{-1})_- L\cr
& - (-)^{| D_{W(E)}|| D_{W(F)}|} (E \lra F)\cr
={}& (L[E,F]L^{-1})_- L + (-)^{n|E|}(L D_{W(E)}FL^{-1})_- L
\cr
&-(-)^{|F|(n+|E|)}(L D_{W(F)}EL^{-1})_- L\cr
={}& W([E,F]_L)~.\()\cr}
$$
\QED

\Rmk<>{ Notice that in the case where $E$ and $F$ are independent of
$L$ we recover the usual Lie bracket on $\SDOP$.}

We have in this moment the following diagram where both maps $W$ and
$J$ have been proven to be Lie algebra homomorphisms:
$$
\commdiag{&&T^*_L\gM\cr
&&\mapdown{J}\cr
\SDOP&\mapright{W}& T_L\gM\cr}$$
It would be thus interesting to see whether one can complete this
diagram with a map $R$ such that $J\circ R = W$ and, if this is
possible, to check whether the new map $R$ is a homomorphism as well.

We consider therefore the map $R : \SDOP\to T_L^*\gM$ defined
by
$$
R(E) = -(EL^{-1})_- \bmod D^{-n}{\cal R}_-~,\()
$$
for any $E$ in $\SDOP$. Since $D^{-n}{\cal R}_-\subseteq\ker J$ we
have that indeed
$$
J\circ R(E) = W(E)~,\()
$$
for any $E$ in $\SDOP$ and therefore $J\circ R=W$.

\Thm<>{$R$ is a Lie algebra homomorphism, with
$$
[R(E),R(F)]_L^* = R([E,F]_L)~.\()
$$}

\Proof Using the fact that $J\circ R=W$ and that $|R(E)|=|E|+n$ we
have
$$
\eqnalign{
[R(E),R(F)]_L^* ={}& -(-)^{n|E|} D_{W(E)}(FL^{-1})_- +
(EL^{-1})_- (L(FL^{-1})_-)_-\cr
&- (((EL^{-1})_- L)_+(FL^{-1})_-)_- - (-)^{|E||F|}(E\lra F)\cr
={}& -(-)^{n|E|}( D_{W(E)}FL^{-1})_- + (-)^{|E||F|}
(FL^{-1}W(E)L^{-1})_-\cr
&+ (EL^{-1})_-(LFL^{-1})_- - (E(FL^{-1})_-)_-\cr
& + ((EL^{-1})_+ L(FL^{-1})_-)_- - (-)^{|E||F|} (E\lra F)\cr
={}& -([E,F]L^{-1})_- - (-)^{n|E|}( D_{W(E)}FL^{-1})_- \cr
&+(-)^{n|F|+|E||F|} ( D_{W(F)}EL^{-1})_-\cr
={}& R([E,F]_L)~.\()\cr}
$$
\QED

\Cor<>{We have the following commutative diagram of Lie algebras:
$$
\def\mapne#1{\llap{$\vcenter{\hbox{$\scriptstyle #1$}}$}\nearrow}
\commdiag{&&T_L^*\gM\cr
&\mapne{R\,}&\mapdown{J}\cr
\SDOP&\mapright{W}& T_L\gM\cr}
$$
}

\section{Additional symmetries of the MRSKP hierarchy}

We shall start in this section the study of the additional symmetries
of supersymmetric KP hierarchies by considering the supersymmetric
extension of KP defined by Manin and Radul in \[MRSKP], the MRSKP
hierarchy.

One of the most remarkable features of the MRSKP hierarchy is the fact
that it possesses a standard Lax formulation, although at the price
that the (odd) flows do not commute with each other.  It is due to its
Lax formulation that this hierarchy has received the most attention
from the physics community.  Indeed, since the KP hierarchy is
connected to $2d$ quantum gravity via the Lax formalism, one might
expect that MRSKP be relevant for $2d$ quantum supergravity.  So far,
however, it seems that the relevant supersymmetric hierarchies in $2d$
quantum supergravity are more or less naive supersymmetrizations of
the KdV-type hierarchies \[Beckers] \[SKdVB] \[SBouss].

In order to study its additional symmetries, let us first briefly
recapitulate a few basic things about MRSKP that will be useful in the
following.  The MRSKP hierarchy is defined as the universal family of
isospectral flows deforming a $\SPDO$ $\Lambda = D +\sum_{i\geq 1}U_i
D^{1-i}$, with $U_i\in R$. But in contrast to the nonsupersymmetric
case this infinite family of odd and even flows satisfy a {\it
nonabelian} Lie superalgebra whose commutation relations are
$$
[D_{2i},D_{2j}]=0~,\quad [D_{2i},D_{2j-1}]=0~,\quad
[D_{2i-1},D_{2j-1}]=-2 D_{2i+2j-2}~.\(MRalgebra)
$$
(We have adopted here the same sign conventions for the time
parameters $t_i$ as in \[MulaseMR], \[Rabin].) A particular
representation of \(MRalgebra) in terms of an infinite number of odd
and even times $\{t_1,t_2,t_3,\ldots\}$ is given by
$$
\eqnalign{
D_{2i}&= \pder{}{t_{2i}}\cr
D_{2i-1}&= \pder{}{t_{2i-1}} - \sum_{j\geq 1}t_{2j-1}
           \pder{}{t_{2i+2j-2}}~,\(MRflows)\cr}
$$
where the odd times are odd variables satisfying
$t_{2i-1}t_{2j-1}=-t_{2j-1}t_{2i-1}$ and $t_{2i-1}^2=0$. These flows
are initially defined on $R$ but one can extend them on the whole
$\cal R$ as evolutionary derivations, that is,
$$
[D_{2i},D]=[D_{2i-1},D]=0~,\(number)
$$
and one can thus write the Lax flows of the MRSKP hierarchy in the
following manner:
$$
\veqnalign{
D_{2i}\Lambda&= -[\Lambda_-^{2i},\Lambda] =
                 [\Lambda_+^{2i},\Lambda]\cr
D_{2i-1}\Lambda&= -[\Lambda_-^{2i-1},\Lambda] =
                   [\Lambda_+^{2i-1},\Lambda] -
                        2\Lambda^{2i}~.\(Laxflows)\cr}
$$
The necessary and sufficient condition for the existence of an even
$\SPDO$, $\phi =1+\sum_{\geq 1}V_iD^{-i}$, with $V_i\in R$, such that
$\Lambda=\phi D\phi^{-1}$---that is, for $\Lambda$ to be
dressable---is in this case $U_1^{[1]}+2U_2=0$.  If we restrict
ourselves to such $\Lambda$'s then we can alternatively define the
MRSKP hierarchy as the family of flows on the dressing operator $\phi$
$$
D_i\phi = -\left(\phi D^i\phi^{-1}\right)_-\phi~,\(dressMR)
$$
or equivalently
$$
\eqnalign{
\pder{\phi}{t_{2i}}&= -\left(\phi D^{2i}\phi^{-1}\right)_-\phi\cr
\pder{\phi}{t_{2i-1}}&= -\left(\phi\left(D^{2i-1}+\sum\nolimits_{j\geq
1}t_{2j-1}D^{2i+2j-2}\right)\phi^{-1}\right)_-\phi~.\()\cr}
$$

\Rmk<>{One can prove that provided $\Lambda$ satisfies the
dressability condition the two definitions of the MRSKP hierarchy are
indeed equivalent.  First of all it is obvious that given the flows on
the dressing operator \(dressMR) one obtains the Lax flows on
$\Lambda$:
$$
\eqnalign{
D_i \Lambda &= D_i\left(\phi D\phi^{-1}\right)\cr
            &= -\left(\phi D^i\phi^{-1}\right)_-\phi D\phi^{-1} +
             (-)^i\phi D\phi^{-1}\left(\phi D^i\phi^{-1}\right)_- \cr
            &= -[\Lambda^i_-,\Lambda]~.\()\cr}
$$
In order to prove the converse let us introduce the dressed version
of $\Lambda$ in the Lax flows and rewrite them in the following form
$$
\left(D_i\phi +\left(\phi D^i\phi^{-1}\right)_-\phi\right)D\phi^{-1} -
(-)^i\phi D\phi^{-1}\left(D_i\phi +\left(\phi
D^i\phi^{-1}\right)_-\phi\right)\phi^{-1} = 0~.\()
$$
Suppose now that $D_i\phi +\left(\phi D^i\phi^{-1}\right)_-\phi
= A_N D^N + A_{N-1}D^{N-1} + \ldots$, for some arbitrary $N$. Then one
obtains the following  conditions for the leading coefficients:
$$
\veqnalign{
A_N=0&\quad\hbox{for } N \hbox{ odd},\cr
\noalign{\hbox{and}}
\left.\matrix{
2A_{N-1}-(-)^n A'_N -(-)^n 2V_1 A_N=0\cr
A'_{N-1}+2V_1 A_{N-1} -(-)^n V'_1 A_N=0}\right\}&\quad\hbox{for }
N\hbox{ even}. \()\cr}
$$
That is, in both cases we obtain that---provided we drop the
constants---the leading coefficient $A_N$ must vanish and hence
\(dressMR) is satisfied.}

\Rmk<>{Notice that one can dress the following obvious commutation
relations
$$
\eqnalign{
[D_{2i}-D^{2i},D]&= 0\()\cr
[D_{2i-1}-D^{2i-1},D]&= -2D^{2i}\(odd)\cr}
$$
and obtain the Lax flows \(Laxflows).}

The flows \(dressMR) are reminiscent of the supersymmetric Radul map
(where in this case $n=1$) and suggest us to define a map $W' : \SDOP
\to {\cal R}_-$ by
$$
W'(E) = \left(\phi E\phi^{-1}\right)_-\phi~,\(W'map)
$$
for any $\SDOP$ $E$ such that $D_n\phi=-W'(D^n)=-D_{W'(D^n)}\phi$,
where $D_{W'(D^n)}$ is a flow on the
Volterra group. The algebra of flows of MRSKP becomes in light of this
definition a simple consequence of \<SRadul>.

\Prop<>{The MRSKP flows  satisfy the Lie superalgebra given in
\(MRalgebra).}

\Proof Following step by step the proof of \<SRadul> and replacing
$L$ with $\phi$ we have that
$$
[D_{W'(E)},D_{W'(F)}] = D_{W'([E,F]_{\phi})}~.\()
$$
Applying this to our MRSKP flows we get for instance
$$
\eqnalign{
[D_{2i-1},D_{2j-1}]&= [D_{W'(D^{2i-1})}, D_{W'(D^{2j-1})}]\cr
                   &= D_{W'(2D^{2i+2j-2})}\cr
                   &= -2 D_{2i+2j-2}~.\()\cr}
$$
One can in a similar way check that all the other commutators in
\(MRalgebra) do indeed vanish.\QED

After these general considerations concerning the MRSKP
hierarchy we are now prepared to tackle the problem of finding its
(additional) symmetries. We have seen that in the case of KP one
defines a larger family of flows (\ie, containing the KP flows) which
satisfy an algebra whose center is the KP hierarchy itself.  Here the
situation will turn out to be slightly different since the MRSKP flows
themselves do not commute with each other but rather they obey the
nonabelian (super)algebra \(MRalgebra).

\Def<MRaddsymdef>{We call (additional) symmetries of the MRSKP
hierarchy the centralizer of the algebra \(MRalgebra) of flows of
MRSKP in the algebra of times-dependent vector fields on $\gM$.}

Analogous to the nonsupersymmetric case, one way to look for additional
symmetries is to look for operators $\Gamma$ satisfying
$$
[D_i -D^i,\Gamma]=0~.\(mr)
$$
The additional flow associated to $\Gamma$ is then obtained via the
supersymmetric Radul map and is given by $D_{W'(\Gamma)}$.

One obvious solution is $\Gamma=\d$, the even derivation on the ring
${\cal R}$ and the generator (via the Radul map) of the even flows of
the hierarchy.  Notice nevertheless that the odd derivation $D$---the
generator of the odd flows---does not obey \(mr) but for even $i$.
One is therefore forced to conclude that only the even flows are
actually symmetries of the hierarchy, this being the most striking
distinction from the nonsupersymmetric case. Apart from this `trivial'
solution, one can of course ask whether there also exist
times-dependent symmetries of the MRSKP hierarchy. The answer to this
question is the object of the following lemma.

\Lem<addsymMR>{Let $R[t_i]$ be the extension ring of $R$ by the time
variables $\{t_i\}$ and let
$$
\eqnalign{
\Gamma_0 &= x + \half \sum_{j\geq 1}jt_j D^{j-2} -\half
\sum_{j\geq1}t_{2j-1}\d^{j-2} Q + \half\sum_{i,j\geq1} (i-j) t_{2i-1}
t_{2j-1} \d^{i+j-2}~,\cr
\Gamma_1 &= \theta + \sum_{j\geq 1}t_{2j-1}\d^{j-1}~,\cr}
$$
where $Q=\d_\theta - \theta\d$, be formal infinite order
(super)differential operators in $R[t_i][[D]]$ of $\Z_2$-degrees
$|\Gamma_0|=0$, $|\Gamma_1|=1$. These operators enjoy the following
properties:\nl
\item{(i)}{$[D_i - D^i,\Gamma_0] = 0$, $[D_i - D^i,\Gamma_1] = 0$, and
$[D_i-D^i, Q]=0$ for any $i\geq 1$;}
\item{(ii)}{$[Q,\Gamma_1] = 1$, $[Q,\Gamma_0]=-\Gamma_1 $, $[\d,
\Gamma_0]=1$;}
\item{(iii)}{$[\Gamma_0,\Gamma_1]=0$, $[\Gamma_1,\Gamma_1]=0$,
      $[\Gamma_0,\Gamma_0]=0$.}
}

\Proof There is one point that ought to be mentioned here, concerning
the extension of the derivations $\d$ and $D$ to the ring $R[t_i]$.
We recall that in the case of the KP hierarchy the first even time
could be identified with $x$ because of the first flow which read
$\d_1=\d$. Here, although the first even flow tells us again that
$D_2=\d$, things turn out to be different. Indeed $D$ cannot be
(analogously to $\d$) identified with  $D_1$, as one can easily
convince oneself by writing down the first odd flow. We are therefore
forced to proceed safely and do not identify $x$ with $t_2$, but
rather keep them as independent variables and extend trivially the
action of $\d$ and $D$ to the ring $R[t_i]$. \QED

We can now define the `additional' flows of the MRSKP hierarchy as the
following four-parameter family of odd and even flows
$$
D_{m,k,l,p}\phi = W'(\Gamma_0^k\Gamma_1^l Q^p \d^m)~,\(rain)
$$
with $k\geq 0$, $l=0,1$, $p=0,1$ and $m\in\Z$, where the even MRSKP
flows can be obtained as a particular case, namely
$D_{m,0,0,0}=-D_{2m}$ for $m>0$.

\Thm<>{The additional flows are symmetries of the MRSKP hierarchy,
that is they commute with the MRSKP flows:
$$
[D_i,D_{m,k,l,p}] = 0~.\()
$$
}

\Proof Using the expression of the flows in terms of the
supersymmetric Radul map and \<SRadul> we have that
$$
[D_i,D_{m,k,l,p}] = D_{W'(-[D^i,\Gamma_0^k\Gamma_1^l
Q^p\d^m]_{\phi})}~,\()
$$
with
$$
\eqnalign{
[D^i,\Gamma_0^k\Gamma_1^l Q^p \d^m]_{\phi} &=
D_{W'(D^i)}\Gamma_0^k\Gamma_1^l Q^p \d^m -
(-)^{i(l+p)}D_{W'(\Gamma_0^k \Gamma_1^l Q^p \d^m)}D^i\cr
&\qquad +[D^i,\Gamma_0^k\Gamma_1^l Q^p \d^m]\cr
&= -[D_i - D^i,\Gamma_0^k\Gamma_1^l Q^p\d^m]~,\()\cr}
$$
which using \<addsymMR> gives us the announced result.\QED

\Cor<>{The algebra of additional symmetries of the MRSKP hierarchy
given by \(rain) is isomorphic to the Lie algebra of SDOP, which is
isomorphic (as a Lie algebra) to $\SW_{1+\infty}$.}

\Proof Indeed, the isomorphism is given by
$$
\eqnalign{
z&\mapsto -\d~,\qquad \xi\mapsto Q+\Gamma_1\d\cr
\d_z&\mapsto \Gamma_0~,\qquad \d_{\xi}\mapsto \Gamma_1~.\()\cr}
$$
The isomorphism between $\SDOP$ and $\SW_{1+\infty}$ is standard (see,
\eg, \[SWinfty]). \QED

\Rmk<>{The fact that we have introduced the generator $Q$ of
supertranslations may seem unsatisfactory to the purist, given that
the MRSKP hierarchy is only defined in terms of abstract derivations
$D_i$ and $D$.  One could therefore ask whether it is really necessary
to break manifest supersymmetric covariance in this fashion instead of
trying to construct another even generator $\widetilde{\Gamma}_0$ that
would behave like $x$ and that would satisfy $[D_i - D^i,
\widetilde{\Gamma}_0] = 0$, $[D,\widetilde{\Gamma}_0]=\Gamma_1$, and
$[\d,\widetilde{\Gamma}_0] = 1$.  This turns out to be impossible,
essentially because $D$ itself is not a symmetry of the hierarchy.
Indeed, an explicit calculation shows that
$$
\eqnalign{
[D_{2i-1} - D^{2i-1},\Gamma_1] &= [D_{2i-1} -
                                  D^{2i-1},[D,{\widetilde \Gamma}_0]]\cr
                               &= -2[\d^i,{\widetilde \Gamma}_0]\cr
                               &= -2i\d^{i-1}~,\()\cr}
$$
which is different from zero and which thus contradicts the theorem.
Hence such an operator ${\widetilde \Gamma_0}$ cannot exist.  One
could nevertheless insist that the very definition of (additional)
symmetry is not appropriate.  Namely, one could argue that by the very
nature of an integrable hierarchy, every flow of MRSKP should be
thought of as a symmetry of all its other flows. In other words one
should include $D$ too as a generator of the additional symmetries.
This would of course require redefining the additional symmetries by
adding to the previously found flows \(rain) the actual flows of the
hierarchy.  One could even go further and claim that once we allowed
for the odd flows of the hierarchy to be part of the additional
symmetries, what we have done is really to relax the condition \(mr)
in order to include \(odd) as a particular case.  But then consistency
would force us to also look for possible times-dependent solutions of
\(mr) where the right hand side would be proportional with an
appropriate power of $\d$.  If one carries on this computation one
finds for instance a whole family of odd operators $\Gamma_1= \theta +
\sum_{j\geq 1}a_j t_{2j-1}\d^{j-1}$ satisfying
$$
\eqnalign{
[D_{2i}-D^{2i},\Gamma_1]&=0~,\()\cr
[D_{2i-1}-D^{2i-1},\Gamma_1]&=(a_i -a_1)\d^{i-1}~.\()\cr}
$$
This embarrassment of riches suggests that this more relaxed notion of
additional `symmetry' is of little interest.}

\section {The \SKP2 Hierarchy}

As it is well known the KP hierarchy can be thought of as a universal
hierarchy for the series of generalized KdV hierarchies, in the sense
that every $n$-th order KdV hierarchy is a reduction of KP, obtained
by imposing $L_-^n=0$. Unfortunately this is no longer true for the
MRSKP hierarchy.  Imposing the constraint $L_-^n=0$ one can still
get integrable hamiltonian hierarchies but not every generalized
$n$-SKdV hierarchy is a reduction of MRSKP.

In a nutshell, this comes about because a superdifferential operator of
order $n$ has a unique $n$-th root if and only if $n$ is odd.
For even $n$, there may not exist an $n$-th root or, even if it
exists, it need not be unique \[MRSKP].  Nevertheless, the fact that
for even $n$ a unique $(\fr n/2)^{\rm nd}$ root does always exist
\[eSKdV] has prompted the study of the so-called even order SKP
hierarchy, \SKP2, and it is the purpose of this section to study its
additional symmetries.

\SKP2 is defined \[eSKdV] as the universal family of isospectral
deformations of a $\SPDO$ of the form
$$
L= D^2 + \sum_{i\geq 1}U_i D^{2-i}~,\()
$$
with $U_i\in R$ and its evolution is described by a {\it commuting\/}
family of flows
$$
\d_i L = -[L_-^i,L] = [L_+^i,L]~,\()
$$
where all the flows are even and therefore can be represented in terms
of an infinite set of even times $\{t_1,t_2,\ldots\}$ by $\d_i=\pder
{}{t_i}$. In the following we shall restrict ourselves to operators
$L$ which are dressable, \ie, which satisfy the conditions $U_1=U_2=0$.

\Rmk<> {Notice that one can dress the following obvious commutation
relations
$$
[\d_n - \d^n, \d] = 0~\()
$$
with an arbitrary $\phi = 1+\sum_{i\geq 1}V_i D^{-i}$, $V_i \in R$,
and obtain the \SKP2 flows.}

Let us now consider the problem of finding the additional symmetries
for this hierarchy. Fortunately we can use our previous experience
with KP and MRSKP to write down the generators of additional
symmetries for \SKP2.  Indeed since the hierarchy has only even flows,
it follows that the $x$-like generator for the additional flows of KP
still commutes with the \SKP2 flows.  Moreover, and because of the
same reason, both $D$ and $Q$ can now be considered generators of
additional symmetries.  In fact we have the following result:

\Lem<addsymSKP2>{Let $R[t_i]$ be the extension ring of $R$ by the even
time variables $\{t_i\}$ and let
$$
\Gamma = x + \sum_{t\geq1} jt_{j} \d^{j-1}
$$
be a formal infinite order (super)differential operator in
$R[t_i][[\d]]$. This operator enjoys the following
properties: $[\d_i - \d^i,\Gamma] = 0$ and $[\d,\Gamma]=1$.  Moreover,
the operators $D$ and $Q$ obey: $[\d_i-\d^i,D]=[\d_i-\d^i,Q]=0$.\QED}

We can now define the `additional' flows of the \SKP2 hierarchy as the
following four-parameter family of odd and even flows
$$
D_{m,k,l,p}\phi = W'(\Gamma^k D^l Q^p \d^m)~,\(raintwo)
$$
with $k\geq 0$, $l=0,1$, $p=0,1$ and $m\in\Z$.  Again, the original
flows of the hirerarchy can be obtained as a particular case, namely
$D_{m,0,0,0}=-\d_{m}$ for $m>0$.

\Thm<>{The additional flows are symmetries of the \SKP2 hierarchy,
that is they commute with the MRSKP flows:
$$
[D_i,D_{m,k,l,p}] = 0~.\()
$$
}

\Proof Using the expression of the flows in terms of the
supersymmetric Radul map and \<SRadul> we have that
$$
[\d_i,D_{m,k,l,p}] = D_{W'(-[\d^i,\Gamma^k D^l Q^p\d^m]_{\phi})}~,\()
$$
with
$$
\eqnalign{
[\d^i,\Gamma^k D^l Q^p \d^m]_{\phi} &=
D_{W'(\d^i)}\Gamma^k D^l Q^p \d^m -
D_{W'(\Gamma^k D^l Q^p \d^m)}\d^i +
[\d^i,\Gamma^k D^l Q^p \d^m]\cr
&= -[\d_i - \d^i,\Gamma^k D^l Q^p\d^m]~,\()\cr}
$$
which using \<addsymSKP2> gives us the announced result.\QED

\Cor<>{The algebra of additional symmetries of the \SKP2 hierarchy
given by \(raintwo) is isomorphic to the Lie algebra of SDOP.}

\Proof Indeed, the isomorphism is given by
$$
\eqnalign{
z&\mapsto -\d~,\qquad \xi\mapsto \half(D+Q) \cr
\d_z&\mapsto \Gamma~,\qquad \d_{\xi}\mapsto \half(D-Q)\d^{-1}~.\()\cr}
$$
\QED

\section{Additional symmetries of the JSKP hierarchy}

Since there is no unique supersymmetric extension of the KP hierarchy
one could of course ask what distinguishes the different
supersymmetric KP hierarchies or which one of them is a more natural
generalization of the KP hierarchy.  We have previously argued that
the MRSKP hierarchy has the advantage of possessing a Lax formulation.
Nevertheless, from a geometrical point of view, it is not the MRSKP
hierarchy the one that seems the most natural supersymmetric
generalization of the KP hierarchy.  Indeed, according to \[Rabin],
one can interpret geometrically the KP flows as deformations of a
certain line bundle over a fixed manifold.  This picture gets slightly
modified in the supersymmetric case where the supersymmetry relation
$D^2=\d$ leads to deformations of the base supermanifold as well as of
the line bundle on it.  From this point of view the MRSKP hierarchy
describes a special subset of deformations in which changes of the
supermanifold are coupled to changes in the line bundle in such a way
that the resulting hierarchy is integrable.  It is the Jacobian SKP
(JSKP) defined by Mulase and Rabin \[MulaseJ]\[Rabin], the one that
seems to provide the the closest geometric analog of the KP hierarchy
in the supersymmetric case since it only involves deformations in the
line bundle.

With this motivation in mind let us define the JSKP hierarchy as the
infinite family of odd and even commuting flows on the Volterra
group given by
$$
\eqnalign{
\pder{\phi}{t_{2i}}&= -\left(\phi\d^i\phi^{-1}\right)_-\phi\cr
\pder{\phi}{t_{2i-1}}&=
-\left(\phi\d^{i-1}\d_{\theta}\phi^{-1}\right)_-\phi~, \()\cr}
$$
where $\phi =1+\sum_{\geq 1}V_iD^{-i}$ and $\{t_1,t_2,t_3,\ldots\}$ is
the same infinite set of odd and even times as in the case of the
MRSKP hierarchy.

\Rmk<>{The even flows of JSKP coincide with the even flows of
MRSKP being actually nothing but the original KP system. On the other
hand, since $\phi\d_{\theta}\not=\d_{\theta}\phi$, it seems there is
no simple way of writing the odd flows in terms of a Lax operator
$L=\phi D\phi^{-1}$. We can nevertheless write the JSKP flows in a Lax
form by defining $L=\phi\d\phi^{-1}$ and $M=\phi\d_{\theta}\phi^{-1}$,
in terms of which the flows can be written as follows:
$$
\pder{L}{t_{2i}} = - [L_-^i,L]~,\qquad \pder{L}{t_{2i-1}} =
-[(L^{i-1}M)_-,L]~.\()
$$
}

Following the same path as for MRSKP it is easily seen that the JSKP
flows can be written in terms of the $W'$ map, \(W'map), by
$$
\eqnalign{
D_{2n}\phi&=-W'(\d^n)=-D_{W'(\d^n)}\phi\cr
D_{2n-1}\phi&=-W'(\d^{n-1}\d_{\theta})= -D_{W'(\d^{n-1}
\d_{\theta})}\phi \()\cr}
$$

\Prop<>{The JSKP flows satisfy a commutative Lie superalgebra.}

\Proof This is already clear since $[\d^n,\d^m] =
[\d^n,\d^m\d_{\theta}] = [\d^n\d_{\theta},\d^m\d_{\theta}] = 0$.\QED

\Lem<asJSKP>{Let $R[t_i]$ be the extension ring of $R$ by the time
variables $\{t_i\}$ and let
$$
\eqnalign{
\Gamma_0 &= x + \sum_{j\geq 1}jt_{2j}\d^{j-1} + \sum_{j\geq 1}
           jt_{2j+1}\d^{j-1}\d_{\theta}\()\cr
\Gamma_1 &= \theta + \sum_{j\geq 1}t_{2j-1}\d^{j-1}\()\cr
\Gamma_2 &= x\d_\theta + \sum_{j\geq
1}jt_{2j}\d^{j-1}\d_{\theta}~.\()\cr}
$$
be formal infinite order differential operators in
$R[t_i][[\d,\d_\theta]]$ of $\Z_2$ -degrees $|\Gamma_0|=0$ and
$|\Gamma_1|=|\Gamma_2|=1$.  These operators have the following
properties:\nl
\item{(i)}{$[D_{2i}-\d^i,\Gamma_k] = 0$ and
$[D_{2i-1}-\d^{i-1}\d_{\theta},\Gamma_k] = 0$ for all $k=0,1,2$;}
\item{(ii)}{$[\d,\Gamma_1]=0$, $[\d_{\theta},\Gamma_1]=1$,
$[\Gamma_1,\Gamma_1]=0$;}
\item{(iii)}{$[\d,\Gamma_2]=\d_{\theta}$, $[\d_{\theta},\Gamma_2]=0$,
$[\Gamma_2,\Gamma_2]=0$;}
\item{(iv)}{$[\d,\Gamma_0]=1$, $[\d_{\theta},\Gamma_0]=0$,
$[\Gamma_1,\Gamma_2]=\Gamma_0$.}
}

\Proof In order to compute the above commutators one has to extend the
action of the derivatives $\d$ and $\d_{\theta}$ to the ring $R[t_i]$.
Here too $\d$ can be identified with the first even flow $D_2$ since
$$
D_2\phi = [\d,\phi]~.\()
$$
If we now consider the first odd flow
$$
D_1\phi = -(\phi\d_{\theta}\phi^{-1})_-\phi~,\(foj)
$$
one may expect that $D_1$ can be identified with $\d_{\theta}$. Still
a short computation of the RHS of \(foj) shows us that this is not
actually the case, but rather $D_1=[\d_{\theta},\phi]+V_1\phi$. It is
therefore safe to extend trivially the action of $\d$ and
$\d_{\theta}$ to $R[t_i]$. \QED

We can now define a three-parameter family of flows
$$
\eqnalign{
D_{2m,k,l}\phi &= W'(\Gamma_0^k\Gamma_1^l\d^m)\cr
D_{2m-1,k,l}\phi &=
           W'(\Gamma_0^k\Gamma_1^l\d^{m-1}\d_{\theta})~,\(addfl)\cr}
$$
where $k\geq 0$, $l=0,1$, and $m\in\integ$. Here the original JSKP
flows are a special case, $D_{m,0,0}=-D_m$ for $m>0$, whereas the
other ones represent the additional symmetries of the JSKP hierarchy.

\Thm<>{The additional flows are symmetries of the JSKP hierarchy, in
other words they commute with the flows on the Volterra group.}

\Proof We only have to use \<SRadul> and we obtain, for instance, for
the even flows
$$
[D_{2i},D_{2m,k,l}]= -
                 D_{W'([\d^i,\Gamma_0^k\Gamma_1^l\d^m]_{\phi})}~,\()
$$
where
$$
\eqnalign{
[\d^i,\Gamma_0^k\Gamma_1^l\d^m]_{\phi} &=
                        D_{W'(\d^i)}\Gamma_0^k\Gamma_1^l\d^m
                              + [\d^i,\Gamma_0^k\Gamma_1^l\d^m]\cr
           &= -[D_{2i}-\d^i,\Gamma_0^k\Gamma_1^l\d^m]\cr
           &= 0~.\()\cr}
$$
Analogous computations give us that
$$
\eqnalign{
[D_{2i},D_{2m-1,k,l}]&= -
D_{W'([\d^i,\Gamma_0^k\Gamma_1^l\d^{m-1}\d_{\theta}]_{\phi})}=0~,\()\cr
[D_{2i-1},D_{2m,k,l}]&= -
D_{W'([\d^{i-1}\d_{\theta},\Gamma_0^k\Gamma_1^l\d^m]_{\phi})}=0~,\()\cr
[D_{2i-1},D_{2m-1,k,l}]&= -
D_{W'([\d^{i-1}\d_{\theta},\Gamma_0^k\Gamma_1^l
\d^{m-1}\d_{\theta}]_{\phi})}=0~,\()\cr}
$$
which finally proves the above statement. \QED

\Cor<>{The Lie superalgebra of symmetries of the Jacobian SKP
hierarchy is isomorphic to $\SDOP$.}

\Proof Let $\cal A$ be the Lie superalgebra of symmetries given by
\(addfl).  It is generated via the Radul map by
$\Gamma_0^k\Gamma_1^l\d^m$ and $\Gamma_0^k \Gamma_1^l \d^{m-1}
\d_{\theta}$ for $k\geq 0$, $l=0,1$ and $m\in\Z$.  The isomorphism
$\SDOP\to{\cal A}$ is given explicitly by
$$
\eqnalign{
z&\mapsto -\d~,\qquad \xi\mapsto \d_{\theta}\cr
\d_z&\mapsto \Gamma_0~,\qquad \d_{\xi}\mapsto \Gamma_1~.\(isom)\cr}
$$
\QED

\Rmk<>{Of the flows $D_{m,k,l}$ defined by \(addfl), all but the
$D_{m,0,0}$ with $m>0$ are additional symmetries.  These additional
symmetries are isomorphic to the direct sum of $\SW_\infty$ with the
abelian algebra generated by the flows $D_{m,0,0}$ with $m\leq 0$.
These flows are present only because the JSKP hierarchy is defined on
the supervolterra group.  If, as in the KP hierarchy, JSKP were
defined on the space of Lax operators, these extra flows would not be
present; for they act trivially on $L= \phi \d \phi^{-1}$.}

The isomorphism of the additional symmetries of all three SKP
hierarchies deserves a final comment. The picture that begins to
emerge is that the additional symmetries, although realized
dinamically with explicit dependence on the times, are actually a
kinematical property of the dynamical system; that is, symmetries of
the phase space in which they are defined.

\ack

I am grateful to J.M.~Figueroa-O'Farrill and E.~Ramos for suggesting
the problem, for many helpful discussions, and for carefully reading
an earlier version of this manuscript.  In addition, I would like to
thank W.~Nahm and V.~Rittenberg for their kind support and
encouragement. Last, but not least, I am thankful to the Physics
Department of Queen Mary and Westfield College for their hospitality
during the final stages of this work.
\refsout
\bye